\input{aipcheck}

\documentclass[
    ,final            
  ]
  {aipproc}

\layoutstyle{6x9}
\def\beq{\begin{equation}}
\def\eeq{\end{equation}}
\newcommand{\MW}{M_W}
\newcommand{\sweff}{\sin^2\theta_{\mathrm{eff}}}
\newcommand{\br}{{\rm BR}}
\def\ga{\mathrel{\raise.3ex\hbox{$>$\kern-.75em\lower1ex\hbox{$\sim$}}}}
\def\la{\mathrel{\raise.3ex\hbox{$<$\kern-.75em\lower1ex\hbox{$\sim$}}}}
\newcommand{\tb}{\tan \beta}

\def\PL{{Phys.~Lett.} }
\def\PR{{Phys.~Rev.} }

\newcommand{\Staue}{\tilde{\tau}_1}
\newcommand{\neu}[1]{\tilde \chi^0_{#1}}
\newcommand{\bmm}{B_s \to \mu^+ \, \mu^-}

\begin{document}
\rightline{hep-ph/0610272}
\rightline{UMN--TH--2518/06}
\rightline{FTPI--MINN--06/31}
\rightline{September 2006}

\title{Searching for Dark Matter in Unification Models: 
A Hint from Indirect Sensitivities towards 
Future Signals in \\
Direct Detection and B-decays\footnote{
Summary of talk given at the International Workshop The Dark Side
Side of the Universe - DSU2006, Univerisdad Autonoma de Madrid, June 2006.}}

\classification{11.30Pb, 13.25.Hw, 95.35.+d, 98.80.-k}
\keywords      {Dark Matter; Supersymmetry}

\author{Keith A. Olive}{
  address={William I Fine Theoretical Physics Institute, \\
University of Minnesota, Minneapolis, MN 55455, USA}
}

\begin{abstract}
 A comparison is made between accelerator and direct detection
 constraints in constrained versions of the minimal supersymmetric
 standard model.   Models considered are based on mSUGRA, 
 where scalar and gaugino masses are unified at the GUT scale.
 In addition, the mSUGRA relation between the (unified) A and B parameters
 is assumed, as is the relation between $m_0$ and the gravitino mass.
 Also considered are models where the latter two conditions are dropped
 (the CMSSM), and a less constrained version where
 the Higgs soft masses are not unified at the GUT scale (the NUHM).

 \end{abstract}

\maketitle


\section{Unification Models}

As constraints from accelerator searches and direct detection experiments
improve, and in anticipation of the potential for discovery at the LHC,
it worthwhile comparing what we can extract from existing data. 
I will assume several unification conditions placed on the
supersymmetric parameters.  In all models considered, the gaugino masses
are assumed to be unified at the GUT scale with value, $m_{1/2}$, 
as are the trilinear couplings with value $A_0$.  Also common to all models
considered here is the unification of all sfermion masses set equal to $m_0$ at the GUT scale.
In the most constrained scenarios \cite{vcmssm}, I will apply the full set of conditions
which are derived in minimal supergravity models (mSUGRA) \cite{bfs}. 
In addition to the conditions listed above,
these include, the unification of all scalar masses (including the Higgs soft masses),
a relation between the bilinear and trilinear couplings $B_0 = A_0 - m_0$, and the
relation between the gravitino mass and soft scalar masses, $m_{3/2} = m_0$.
When electroweak symmetry breaking boundary conditions are applied,
this theory contains only $m_{1/2}, m_0$, and $A_0$ in addition to the sign of the 
Higgs mixing mass, $\mu$, as free parameters. The magnitude of $\mu$ as well
as $\tan \beta$ are predicted.   

The extensively studied \cite{cmssmnew} constrained version of the MSSM or CMSSM drops the
latter two conditions.  Namely, $B_0$ and the gravitino mass are not fixed 
by other parameters.  As a result, $\tan \beta$ becomes a free parameter (as does the
gravitino mass). 
Finally, I will also discuss a less constrained model, the NUHM, in which the Higgs soft
masses are not unified at the GUT scale \cite{nonu,nuhm}.  In this class of models,
both $\mu$ and the Higgs pseudo scalar mass become free parameters.

\section{Indirect sensitivities}

Measurements at low energies can provide interesting indirect information
about the supersymmetric parameter space.
For example, data obtained at the Brookhaven $g_\mu - 2$ experiment \cite{newBNL}
favored distinct regions of parameter space \cite{g-2}. 
Present
data on observables such as $\MW$, $\sweff$, and
$\br(b \to s \gamma)$ in addition to $(g_\mu-2)$ already provide interesting information on the scale of
supersymmetry (SUSY)~\cite{ehow3,ehow4}.
The non-discovery of charginos and the Higgs boson at 
LEP also  imposes significant lower bounds on
$m_{1/2}$. 

An important further constraint is provided by the
density of dark matter in the Universe, which is tightly constrained by
the three-year data from WMAP \cite{wmap} which
has determined many cosmological parameters to
unprecedented precision.  
In the context of the $\Lambda$CDM model, the WMAP only results indicate
\beq
\Omega_m h^2 = 0.1268^{+0.0072}_{-0.0095} \qquad \Omega_b h^2 = 0.02233^{+0.00072}_{-0.00091}
\eeq
The difference corresponds to the requisite dark matter density
\beq
\Omega_{CDM} h^2 = 0.1045^{+0.0072}_{-0.0095}
\eeq
or a 2$\sigma$ range of 0.0855 -- 0.1189 for $\Omega_{CDM} h^2$.

\begin{figure}
\includegraphics[height=2.6in]{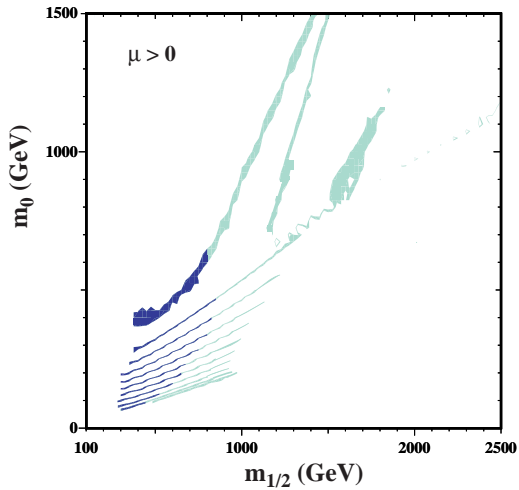}
\caption{\label{fig:strips}\it
The strips display the regions of the $(m_{1/2}, m_0)$ plane that are
compatible with $0.094 < \Omega_\chi h^2 < 0.129$ and the laboratory
constraints for $\mu > 0$ and $\tan \beta = 5, 10, 15, 20, 25, 30,
35, 40, 45, 50, 55$. The parts of the strips compatible with $g_\mu - 2$ 
at the 2-$\sigma$ level have darker shading.
}
\end{figure}

The dark matter constraint has the effect within the CMSSM, assuming that the dark matter 
consists largely of neutralinos~\cite{EHNOS},
of restricting $m_0$ to very narrow allowed strips
for any specific choice of $A_0$, $\tb$ and the sign of 
$\mu$~\cite{eoss,Baer}. These strips are typically due to 
co-annihilation processes between the neutralino and stau \cite{stauco}.
Shown in Fig.~\ref{fig:strips} are the WMAP lines \cite{eoss} of the $(m_{1/2}, m_0)$
plane allowed by the cosmological constraint 
 and laboratory constraints for $\mu > 0$ and
values of $\tan \beta$ from 5 to 55, in steps $\Delta ( \tan \beta ) = 5$.
We notice immediately that the strips are considerably narrower than the
spacing between them, though any intermediate point in the $(m_{1/2},
m_0)$ plane would be compatible with some intermediate value of $\tan
\beta$. The right (left) ends of the strips correspond to the maximal
(minimal) allowed values of $m_{1/2}$ and hence $m_\chi$. 
The lower bounds on $m_{1/2}$ are due to the Higgs 
mass constraint for $\tan \beta \le 23$, but are determined by the $b \to 
s \gamma$ constraint for higher values of $\tan \beta$. 
Thus, the dimensionality of the supersymmetric parameter space is further
reduced, and one may explore supersymmetric phenomenology along these
`WMAP strips', as has already been done for the direct detection of
supersymmetric particles at the LHC and linear colliders of varying 
energies~\cite{bench,otherAnalyses}.

Another
mechanism for extending the allowed CMSSM region to large
$m_\chi$ is rapid annihilation via a direct-channel pole when $m_\chi
\sim {1\over 2} m_{A}$~\cite{funnel,efgosi}. Since the heavy scalar and
pseudoscalar Higgs masses decrease as  
$\tan \beta$ increases, eventually  $ 2 m_\chi \simeq  m_A$ yielding a
`funnel' extending to large
$m_{1/2}$ and
$m_0$ at large
$\tan\beta$, as seen in the high $\tan \beta$ strips of Fig.~\ref{fig:strips}.

For large values of $\tan \beta$, particularly when $m_A$ is small, 
supersymmetry leads to an enhancement in an otherwise rare decay 
of the $B$ meson, namely $\bmm$.  The decay $\bmm$ is known to impose another interesting
constraint on the parameter spaces of models for physics beyond
the Standard Model, such as the MSSM~\cite{Dedes,bmm,bmm2}. 
The Fermilab Tevatron collider already has an interesting upper limit 
$\sim 2 \times 10^{-7}$ on the $\bmm$ decay branching ratio~\cite{cdf}, and
future runs of the Fermilab Tevatron collider and the LHC are expected to 
increase significantly the experimental sensitivity to $\bmm$ decay.
Indeed, the latest CDF limit \cite{cdfnew}, is now $1  \times 10^{-7}$ at the 
95\% CL.
Currently, $\bmm$ does not provide strong constraints in the CMSSM \cite{bmm}.
However, as we will see, in the NUHM, current experimental limits already
exclude interesting models \cite{bmm2}. 

Finally, there is one additional region of acceptable relic density known as the
focus-point region \cite{fp}, which is found
at very high values of $m_0$. 
As $m_0$ is increased, the solution for $\mu$ at low energies as determined
by the electroweak symmetry breaking conditions eventually begins to drop. 
When $\mu \la m_{1/2}$, the composition of the LSP gains a strong Higgsino
component and as such the relic density begins to drop precipitously.  
As $m_0$ is increased further, there are no longer any solutions for $\mu$.

In \cite{ehow3,ehow4} we considered the following observables: the $W$~boson mass, $\MW$, the
effective weak mixing angle at the $Z$~boson resonance, $\sweff$, the
anomalous magnetic moment of the muon, \mbox{$(g_\mu-2)$} (we used the 
SM prediction based on the $e^+e^-$ data for the hadronic vacuum
polarization contribution \cite{Davier})
and the rare
$b$ decays $\br(b \to s \gamma)$, as well
as the mass of 
the lightest $CP$-even Higgs boson, $m_h$. 

We performed the analysis of the sensitivity to $m_{1/2}$ moving 
along the WMAP
strips with fixed values of $A_0$ and $\tb$. The experimental central
values, the present experimental errors and theoretical uncertainties
are as described in \cite{ehow3,ehow4}.
Assuming that the five observables are
uncorrelated, a $\chi^2$ fit has been performed with
\beq
\chi^2 \equiv \sum_{n=1}^{4} \left( \frac{R_n^{\rm exp} - R_n^{\rm theo}}
                               {\sigma_n}  \right) + \chi^2_{m_h}
\eeq
Here $R_n^{\rm exp}$ denotes the experimental central value of the
$n$th observable,
$R_n^{\rm theo}$ is the corresponding theoretical prediction, $\sigma_n$
denotes the combined error, and and $\chi^2_{m_h}$ denotes the
$\chi^2$ contribution coming from the lightest MSSM Higgs boson
mass~\cite{ehow4}.

Our final analysis ingredient is the elastic scattering cross section between
a neutralino and the proton, which is tested by direct detection experiments. 
The following low-energy effective
four-fermion
Lagrangian describes spin-independent elastic $\chi$-nucleon   
scattering:
\begin{equation}
{\cal L} \, = \, \alpha_{3i} \bar{\chi} \chi \bar{q_{i}} q_{i},
\label{lagr}
\end{equation}
which is to be summed over the quark flavours $q$, and the
subscript $i$ labels up-type quarks ($i=1$) and down-type quarks
($i=2$). Expressions for $\alpha_{3i}$ can be found in \cite{eoss8}.
The scalar part of the
cross section can be written as
\begin{equation}
\sigma_{3} = \frac{4 m_{r}^{2}}{\pi} \left[ Z f_{p} + (A-Z) f_{n}
\right]^{2} ,
\label{si}
\end{equation}
where $m_r$ is the reduced LSP mass,
\begin{equation}
\frac{f_{p}}{m_{p}} = \sum_{q=u,d,s} f_{Tq}^{(p)}
\frac{\alpha_{3q}}{m_{q}} +
\frac{2}{27} f_{TG}^{(p)} \sum_{c,b,t} \frac{\alpha_{3q}}{m_q},
\label{f}
\end{equation}
the parameters $f_{Tq}^{(p)}$  are defined by
\begin{equation}
m_p f_{Tq}^{(p)} \equiv \langle p | m_{q} \bar{q} q | p \rangle
\equiv m_q B_q ,
\label{defbq}
\eeq
$f_{TG}^{(p)} = 1 - \sum_{q=u,d,s} f_{Tq}^{(p)} $~\cite{SVZ},
and $f_{n}$ has a similar expression.  
This may be determined from the $\pi$-nucleon $\Sigma$ term, which is 
given by
\beq
\sigma_{\pi N} \equiv \Sigma = {1 \over 2} (m_u + m_d) (B_u + B_d) .
\eeq
and carries substantial uncertainties \cite{eoss8}.
Here we will consider $\Sigma = 45$ and 64 GeV.

\section{mSUGRA models}

We begin the discussion of unifications models, with the most constrained
version of the MSSM, based on mSUGRA, labelled here as the
VCMSSM  \cite{vcmssm}.  Recall that in these models, $\tan \beta$ is fixed by the 
electroweak boundary conditions, and values of $\tan \beta$ are generally
below 35 for most of the $m_{1/2}, m_0$ planes which are now
characterized by $A_0/m_0$. As a result, relic density funnels do not appear,
nor is the focus point ever reached.  The signal from $\bmm$ is very weak and 
direct detection experiments are only beginning to sample these models
(see below for the CMSSM).

Fig.~\ref{fig:VCMSSMGDM} displays the $\chi^2$ function for a sampling of
gravitino dark matter (GDM)  scenarios \cite{gdm} 
obtained by applying the supplementary gravitino mass
condition to VCMSSM models for $A_0/m_0 = 0, 0.75, 3 - \sqrt{3}$ and 2,
and scanning the portions of the $m_{1/2}, m_0$ planes with GDM  at low $m_0$ \cite{ehow4}. 
These wedges are scanned via a
series of points at fixed (small) $m_0$ and increasing $m_{1/2}$. 
As seen in Fig.~\ref{fig:VCMSSMGDM}, the global minimum of $\chi^2$ for
all the VCMSSM models with gravitino dark matter (GDM) 
with $A_0/m_0 = 0, 0.75, 3 - \sqrt{3}$ and 2 is
at $m_{1/2} \sim 450$ GeV. 
As a consequence, there are good prospects for observing the gluino
and perhaps the stop at the LHC. We recall that, in these GDM scenarios, the
$\Staue$ is the NLSP, and that the $\neu{1}$ is heavier. The
$\Staue$ decays into the gravitino and a $\tau$, and is
metastable with a lifetime that may be measured in hours, days or weeks.
Specialized detection strategies for the LHC were discussed
in~\cite{Moortgat}: this scenario would offer exciting possibilities near
the $\Staue$ pair-production threshold at the ILC.

The results indicate that, already at the present level of experimental
accuracies, the electroweak precision observables combined with the WMAP
constraint provide a sensitive probe of the VCMSSM, yielding interesting
information about its parameter space. The rise in $\chi^2$ at 
low $m_{1/2}$ is primarily due to the constraint
from $m_h$, whereas at high $m_{1/2}$, the rise is due to the discrepancy 
between the $g_\mu - 2$ measurement and the standard model calculation. 
Also important however are the contributions from $\MW$ and $\sweff$
as will be seen in the analogous discussion for the VCMSSM.

\begin{figure}[htb!]
\includegraphics[width=.48\textwidth]{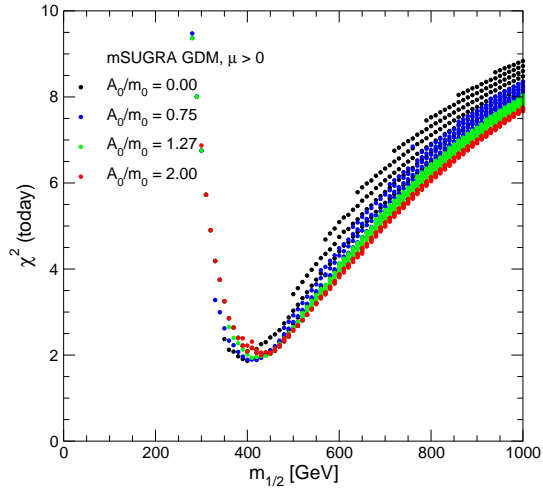}
\caption{\it
The dependence of the $\chi^2$ function on $m_{1/2}$ for
GDM scenarios with $A_0/m_0 = 0, 0.75, 3 - \sqrt{3}$ and $2$,
scanning the regions where the lighter stau $\Staue$ is the
NLSP.
} 
\label{fig:VCMSSMGDM}
\end{figure}

As discussed above, a feature of the class of GDM scenarios discussed here
is that the required value of $\tb$ increases with $m_{1/2}$.  Therefore,
the preference for relatively small $m_{1/2}$ discussed above maps into an
analogous preference for moderate $\tb$.
 We found that, at the 95\% confidence level
\begin{equation}
300 {\rm GeV} \la m_{1/2} \la 800 {\rm GeV}, \qquad 15 \la \tb \la 27
\label{GDMlimits}
\end{equation}
in this mSUGRA class of GDM models. 

\section{The CMSSM}

When we drop the conditions on $B_0$ and $m_{3/2}$,
we recover the well studied CMSSM.  $\tan \beta$ is now a 
free parameter, and we will assume that the gravitino is 
suitably heavy so as to allow for neutralino dark matter.
For a given value of $\tan \beta$ and $A_0$, the relic density 
can be used to fix $m_0$ as a function of $m_{1/2}$ producing the
WMAP strips seen in Fig.~\ref{fig:strips}. 
The first panel of Fig.~\ref{fig:newmt10}, 
displays the behaviour of the $\chi^2$ function out to the tips
of typical WMAP coannihilation strips.  As one can see,
there is a pronounced minimum in $\chi^2$ as a function of 
$m_{1/2}$ for $\tb = 10$. The $\chi^2$ curve depends strongly on the value 
of $A_0$, corresponding to its strong impact on $m_h$. 
Values of $A_0/m_{1/2} < -1$ are disfavoured at the 90\%~C.L.,
essentially because of their lower $m_h$ values, but
$A_0/m_{1/2} = 2$ and~1 give equally good fits and descriptions of the
data. The constraint due to $m_h$ is chiefly responsible for the sharp
rise in $\chi^2$ at low $m_{1/2}$.

\begin{figure}[htb!]
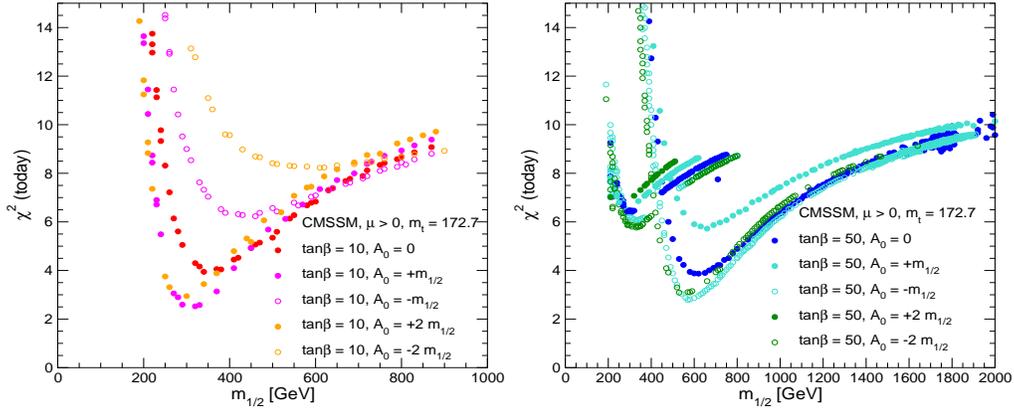

\includegraphics[width=.45\textwidth,height=5.4cm]{ehow.CHI11a.1727.cl.eps}
\includegraphics[width=.45\textwidth,height=5.4cm]{ehow.CHI11b2.1727.cl.eps}
\caption{
\it The combined likelihood function $\chi^2$ for the electroweak
observables $\MW$, $\sweff$,
$(g - 2)_\mu$, ${\rm BR}(b \to s \gamma)$, and $m_h$
evaluated in the CMSSM for $\tb = 10$ (a) and 50 (b),
$m_t = 172.7 \pm 2.9$ GeV and various discrete values of $A_0$, with
$m_0$ then chosen to yield the central value of the relic neutralino 
density
indicated by WMAP and other observations. } 
\label{fig:newmt10}
\end{figure}

At large $m_{1/2}$, the increase in $\chi^2$ is largely due to $g_\mu - 2$,
but also has sizable contributions from the observables $\MW$ and $\sweff$.
The importance of these latter two observables
has grown with recent determinations of $m_t$.
The previous range $m_t = 178.0 \pm 4.3$ GeV~\cite{oldmt}
has evolved to 
$172.7 \pm 2.9 $ GeV~\cite{newmt} (and very recently to 
$172.5 \pm 2.3 $ GeV ~\cite{newestmt} and even more recently to 
$171.4 \pm 2.3$ GeV ~\cite{evennewstmt}). 
The effect of this lower $m_t$ value is twofold \cite{ehow4p}.

First, it drives the
SM prediction of $\MW$ and $\sweff$ further away from the
current experimental value~\footnote{Whereas $(g-2)_\mu$ and 
$\br(b \to s \gamma)$ are 
little affected.}. This effect is shown in Fig.~\ref{fig:MW}
for $\tb = 10$.
The change in the SM prediction elevates
the experimental discrepancy to about 1.5 $\sigma$, despite the change
in the preferred experimental range of $\MW$, which does not
compensate completely for the change in $m_t$. The net effect is
therefore to increase the favoured magnitude of the supersymmetric
contribution, i.e., to lower the preferred supersymmetric mass scale.
In the case of $\sweff$, the reduction in $m_t$ has increased the SM
prediction whereas the experimental value has not changed
significantly. Once again, the discrepancy with the SM has increased
to about 1.5 $\sigma$, and the preference for a small value of
$m_{1/2}$ has therefore also increased. 
With the
new lower experimental value of 
$m_t$, $\MW$ and $\sweff$ give substantial contributions, adding up 
to more than 50\% of the $(g-2)_\mu$ contribution to $\chi^2$ at the tip of the WMAP strip.
Secondly, the predicted value of the lightest Higgs boson mass in the
MSSM is lowered by the new $m_t$ value.
As a result, the LEP Higgs bounds~\cite{LEPHiggsSM} now impose a more
important constraint on the MSSM parameter space, notably on $m_{1/2}$.

\begin{figure}[htb!]
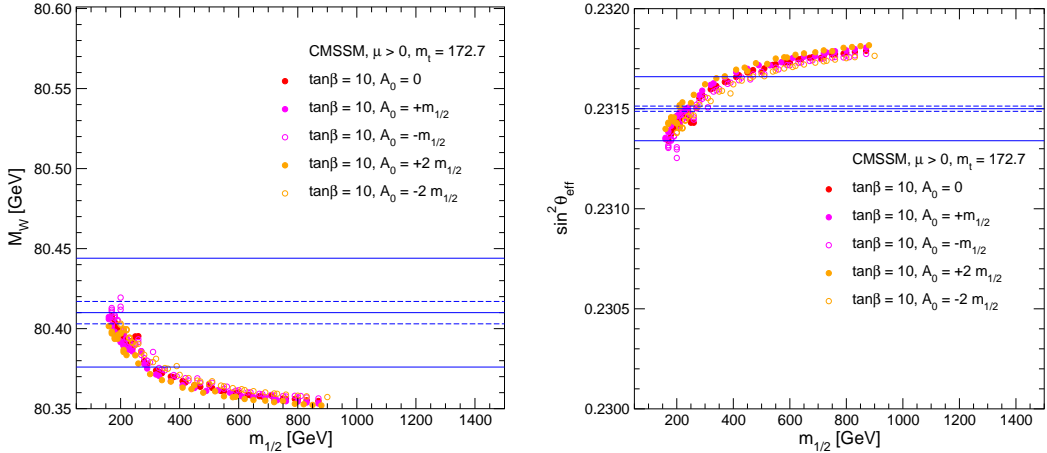

\includegraphics[width=.45\textwidth,height=6cm]{ehow.MW11a.1727.cl.eps}\hspace{1em}
\includegraphics[width=.45\textwidth,height=6cm]{ehow.SW11a.1727.cl.eps}
\vspace{-0.5cm}
\caption{
\it The CMSSM predictions for $\MW$ (a) and $\sweff$ (b)  as functions of $m_{1/2}$ for 
$\tb = 10$ for various $A_0$. The top quark mass has been set to 
$m_t = 172.7$ GeV. The current
experimental measurements indicated in the plots are shown by the solid lines.
Future ILC sensitivities are estimated by the dashed lines.} 
\label{fig:MW}
\end{figure}

The corresponding results for 
WMAP strips in the coannihilation, Higgs funnel and focus-point regions for
the case $\tb = 50$ are shown in
Fig.~\ref{fig:newmt10}b.  The spread of points with identical values of
$A_0$ at large $m_{1/2}$ is due to the broadening and bifurcation of
the WMAP strip in the Higgs funnel region. With the lower value of $m_t$,
there is the
appearance of a group of points with 
moderately high $\chi^2$ that have relatively small $m_{1/2} \sim 200$ GeV.
These points have relatively large values of $m_0$ and are
located in the focus-point region of the $(m_{1/2}, m_0)$
plane~\cite{fp}. 
By comparison with our previous analysis, the focus-point region appears at
considerably lower values of $m_0$, because of the reduction in the central
value of $m_t$. This focus-point strip extends to larger values of
$m_0$ and hence $m_{1/2}$ that are not shown. 
The least-disfavoured focus points have a 
$\Delta \chi^2$ of at least~3.3, and most of them are excluded at the 90\%~C.L. 

Taken at face value, the preferred ranges for the sparticle masses shown
in Fig.~\ref{fig:newmt10}  are quite
encouraging for both the LHC and the ILC. The gluino and squarks lie
comfortably within the early LHC discovery range, and several
electroweakly-interacting sparticles would be accessible to ILC(500)
(the ILC running at $\sqrt{s} = 500$ GeV). The
best-fit CMSSM point is quite similar to the benchmark point
SPS1a \cite{sps} (which is close to point benchmark point B \cite{bench})
which has been shown to offer good
experimental prospects for both the LHC and ILC~\cite{lhcilc}. 
The prospects for sparticle detection are also quite good in the
least-disfavoured part of the focus-point region for $\tb = 50$ shown
in Fig.~\ref{fig:newmt10}b, with the exception of the relatively heavy
squarks.

Direct detection techniques
rely on an ample neutralino-nucleon scattering cross-section.
In Fig.~\ref{fig:EHOW3}a, we display the expected ranges of 
the spin-independent  cross sections in the CMSSM when we
sample randomly $\tan \beta$ as well as the other CMSSM parameters \cite{eoss8}. 
Also shown on the plot is the current CDMS \cite{cdms} exclusion curve
which places an upper limit on the scattering cross section.  As one can 
see, the current limits have only just now begun to probe
CMSSM models.  CMSSM parameter choices with
low $\chi^2$ based on the indirect sensitivities discussed above are
shown in panel b where both the 68\% and 95\% CL points for $\tan \beta = 10$ and 50 are displayed.
These points remain below current the CDMS sensitivity.

\begin{figure}
\includegraphics[width=.45\textwidth,height=6cm]{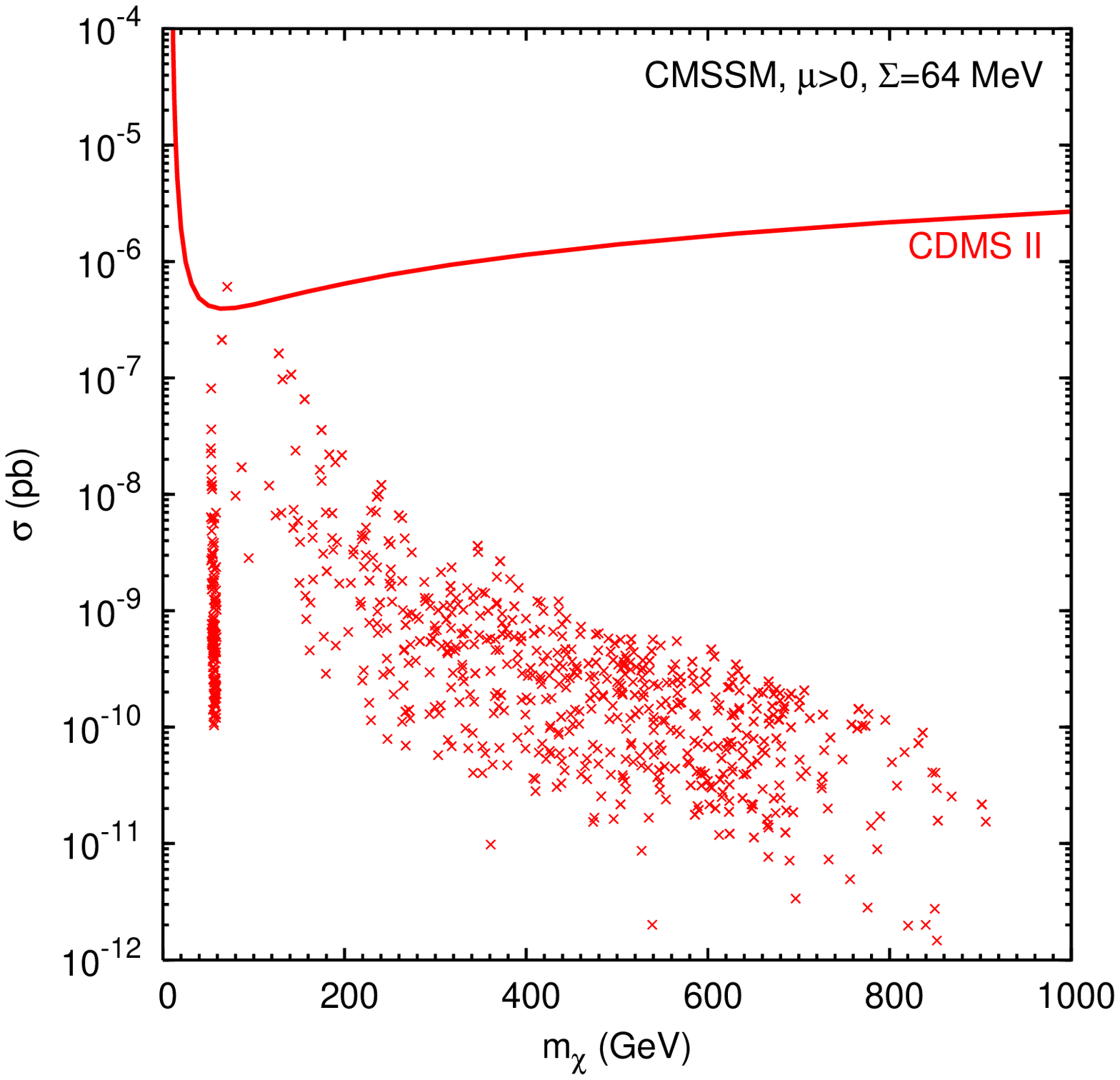}
\includegraphics[width=.45\textwidth,height=6cm]{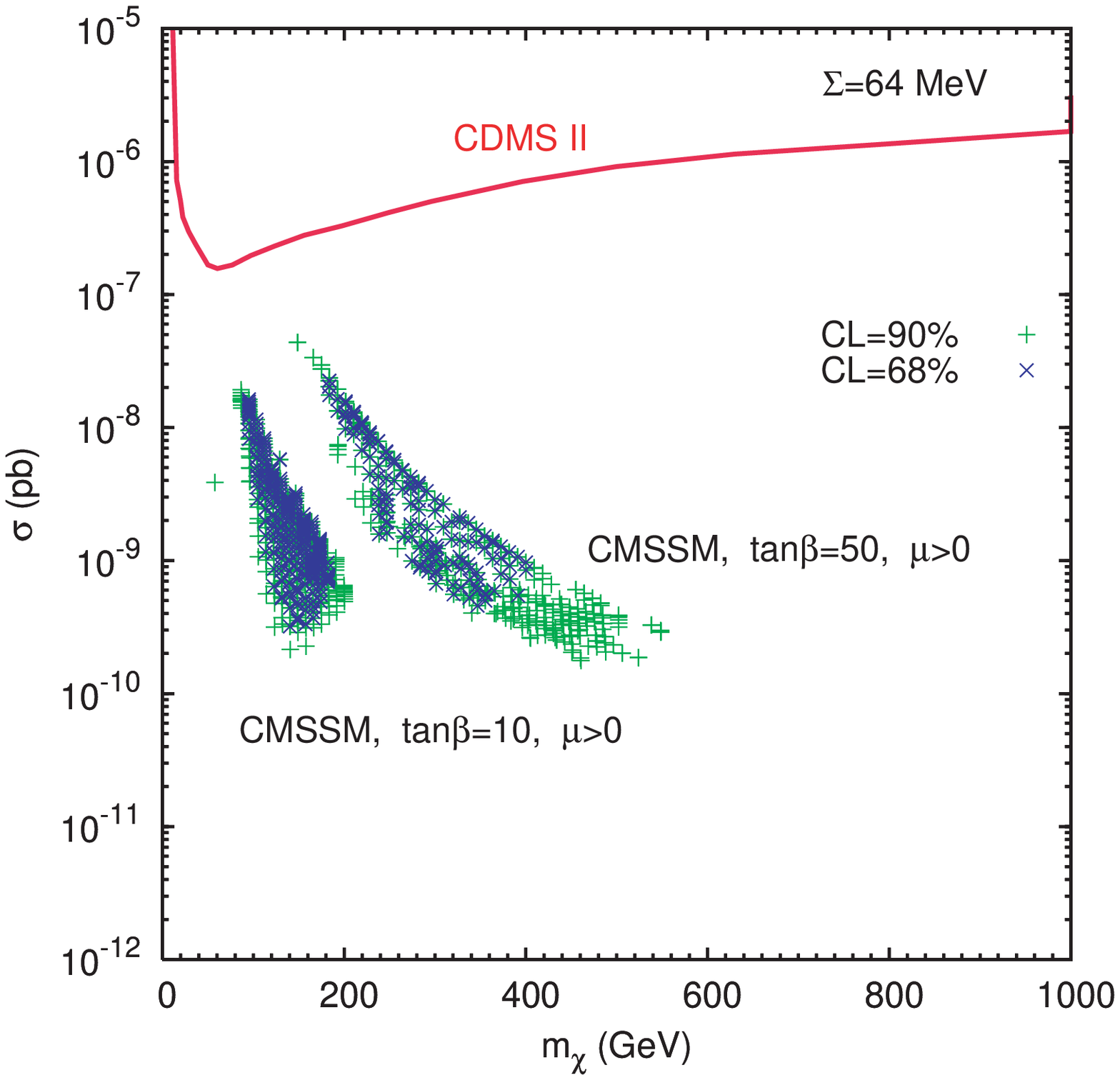}
\caption{\label{fig:EHOW3}
{\it Scatter plots of the spin-independent elastic-scattering cross
section predicted in the CMSSM for  $\tan \beta = 10, \mu > 0$, with 
$\Sigma = 64$~MeV. In panel b, the predictions for models allowed at the 68\% (90\%) confidence 
levels are shown by blue $\times$ signs (green $+$ signs).}} 
\end{figure}

\section{NUHM Models}

In the NUHM, we can either choose the two Higgs soft masses as additional free parameters
or more conveniently we can choose $\mu$ and $m_A$. The addition of new
parameters opens up many possible parameter planes to study.  In \cite{ehow4},
in addition to $m_{1/2}, m_0$ planes with non-CMSSM  values of 
$\mu$ and $m_A$, $\chi^2$'s were computed for $m_{1/2}, \mu$ and $\mu, m_A$
planes. It was concluded that although the preferred value 
of the overall sparticle mass scale set by $m_{1/2}$ may be quite similar 
in the NUHM to its CMSSM value, the masses of some sparticles in the
NUHM may differ significantly from the corresponding CMSSM values.

The NUHM also allows for the possibility of significantly higher
elastic cross sections for $\chi - p$ scattering and 
current constraints already exclude many interesting models \cite{eoss8}.
Furthermore, in those NUHM models
which have cross sections in excess of the CDMS limit, one finds
relatively low values of $\mu$ and $m_A$.

In Fig.~\ref{fig:nuhm}, the current CDMS limit is shown compared
with a scan over the NUHM parameter space.  
In panel a, the value of the $\pi$-nucleon $\Sigma$ parameter was taken as 45 MeV
and can be compared to panel b, where $\Sigma$ = 64 MeV.  The
latter clearly shows higher elastic cross sections and represents an inherent uncertainty
in the theoretical predictions for these cross sections.

\begin{figure}
\includegraphics[width=.45\textwidth,height=6cm]{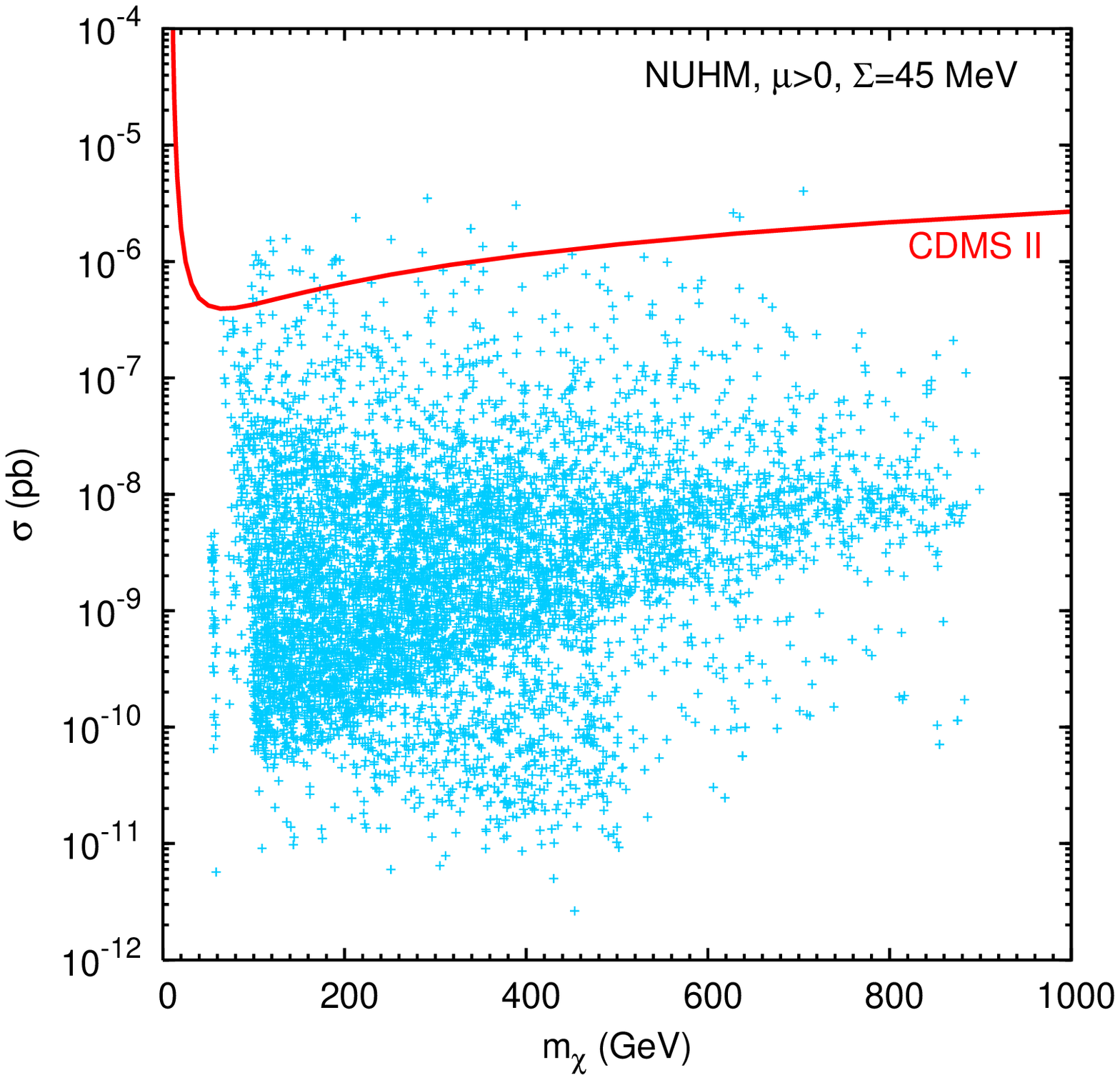}
\includegraphics[width=.45\textwidth,height=6cm]{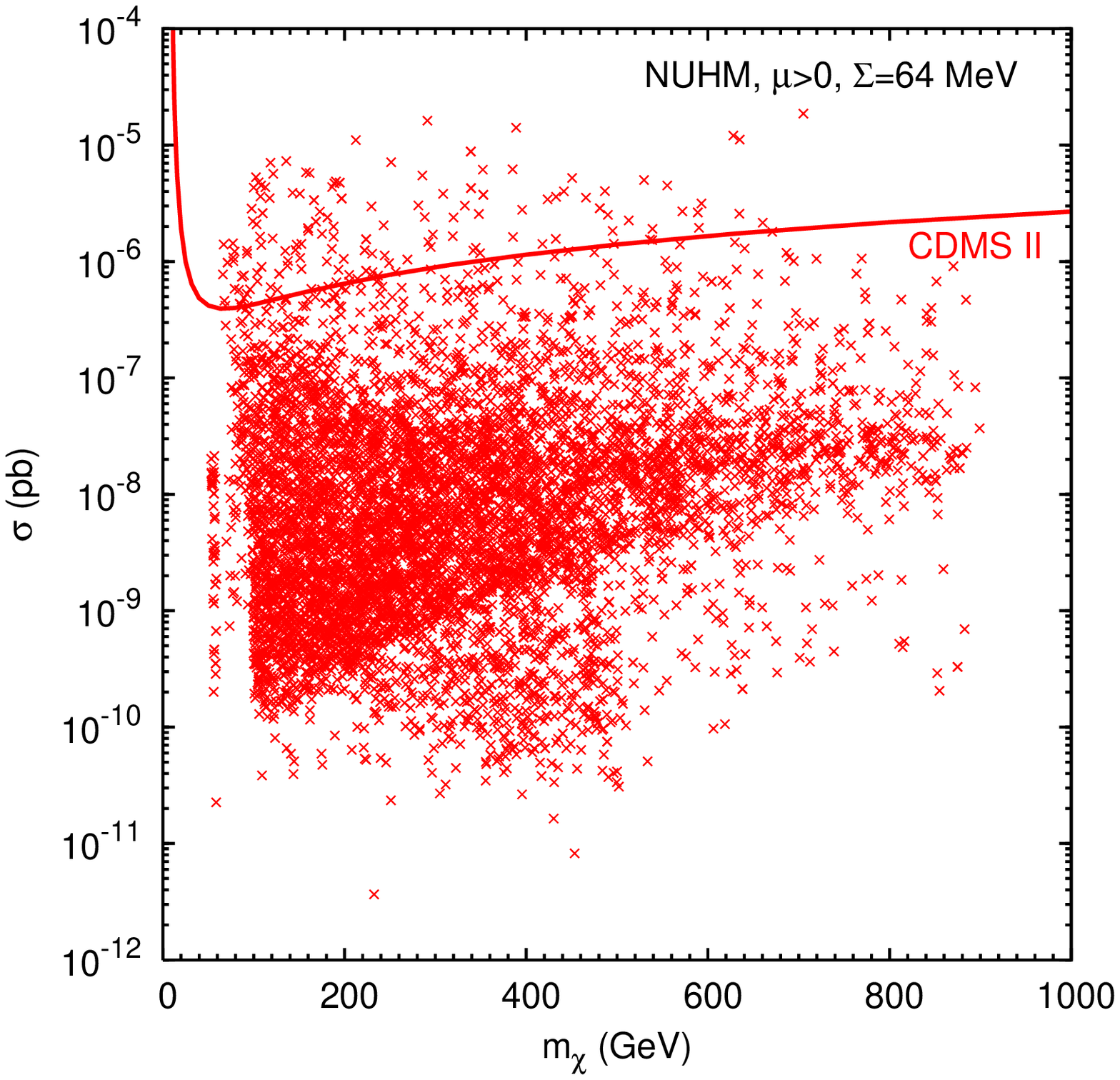}
\caption{\label{fig:nuhm}
{\it Scatter plots of the spin-independent elastic-scattering cross
section predicted in the CMSSM for (a, b) $\tan \beta = 10, \mu > 0$ and 
(c, d) $\tan \beta = 50, \mu > 0$, with (a, c) $\Sigma = 45$~MeV and (b, d) $\Sigma = 
64$~MeV. The predictions for models allowed at the 68\% (90\%) confidence 
levels are shown by blue $\times$ signs (green $+$ signs).}} 
\end{figure}

Some specific NUHM $(\mu, m_A)$ planes for
different values of $\tan \beta$, $m_{1/2}$ and $m_0$ are shown in Fig.~\ref{fig:bmm}, 
exhibiting the interplay of the different experimental, phenomenological and theoretical
constraints \cite{bmm2}. Each panel features a pair
of WMAP strips, above and below the $m_A = 2 m_\chi$ solid (blue) line. 
The WMAP strip is also seen to follow the brick-red shaded region
where the ${\tilde \tau}_1$ is the LSP. 
The lower parts of the 
WMAP strips are excluded by $\bmm$ as $\tan \beta$ increases
as seen by the thick black curve which represents the tevetron limit.
Here one sees clearly the consequence of the improvement to the CDF
bound from 2 to 1 $\times 10^{-7}$ as more of the WMAP strip is now excluded.  
However, in
each case sensitivity to $\bmm$ below $10^{-8}$ would be required to
explore all of the upper WMAP strip.

Also shown in Fig.~\ref{fig:bmm},  is a dashed grey line which is the constraint imposed by
the CDMS upper limit on spin-independent elastic cold dark matter
scattering.
It is interesting to note that the CDMS bound
excludes
a somewhat larger part of the lower WMAP strip than does $\bmm$ for $\tan
\beta = 40$, whereas the $\bmm$ constraint is stronger for $\tan \beta =
50$. Most interesting is the result that these two observables are in fact quite comparable and
one can infer from these figures, that the positive detection of either $\bmm$ or the direct
detection of dark matter should be matched quickly by the detection of the other.

\begin{figure}
\includegraphics[width=.45\textwidth,height=6cm]{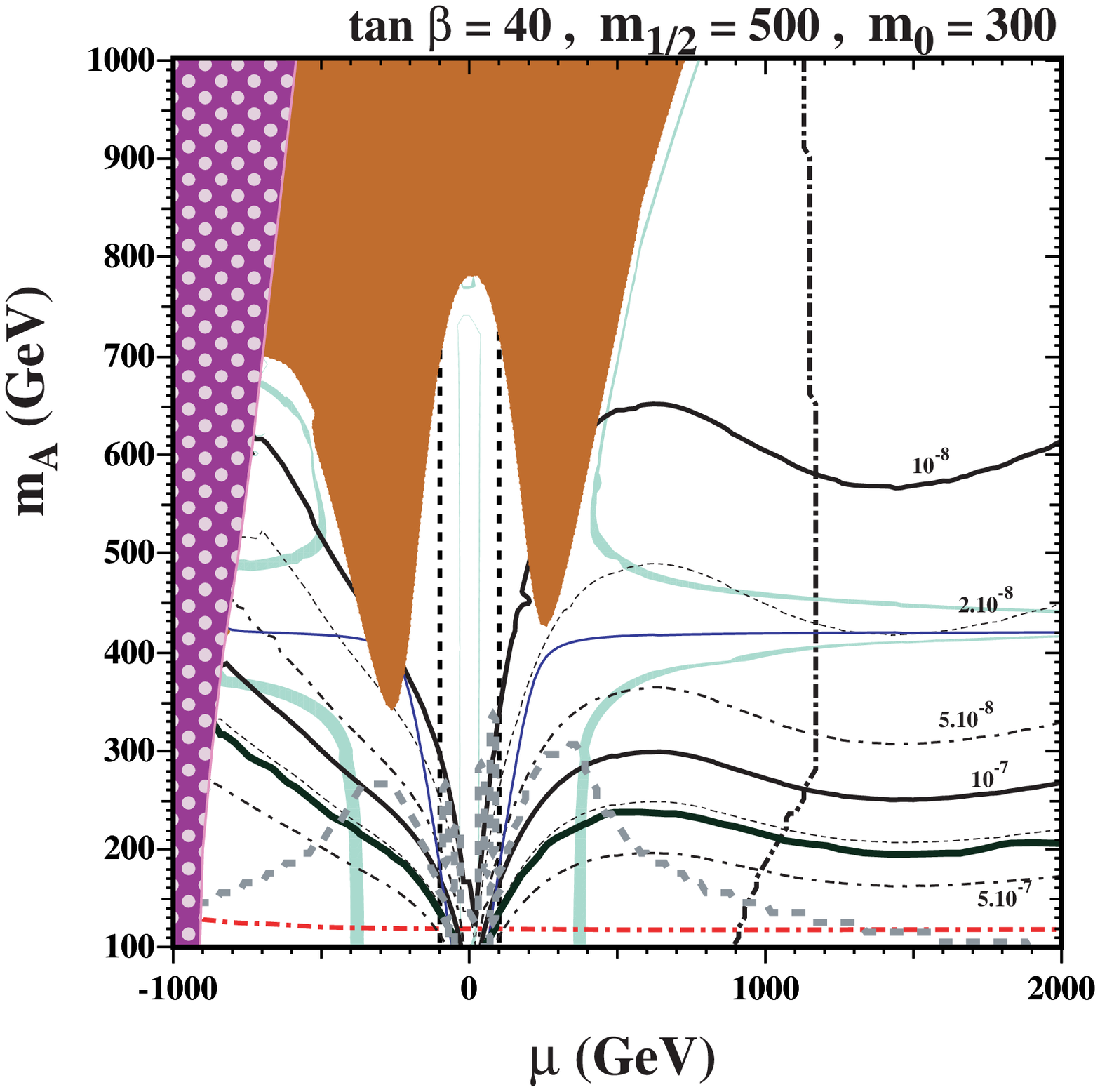}
\includegraphics[width=.45\textwidth,height=6cm]{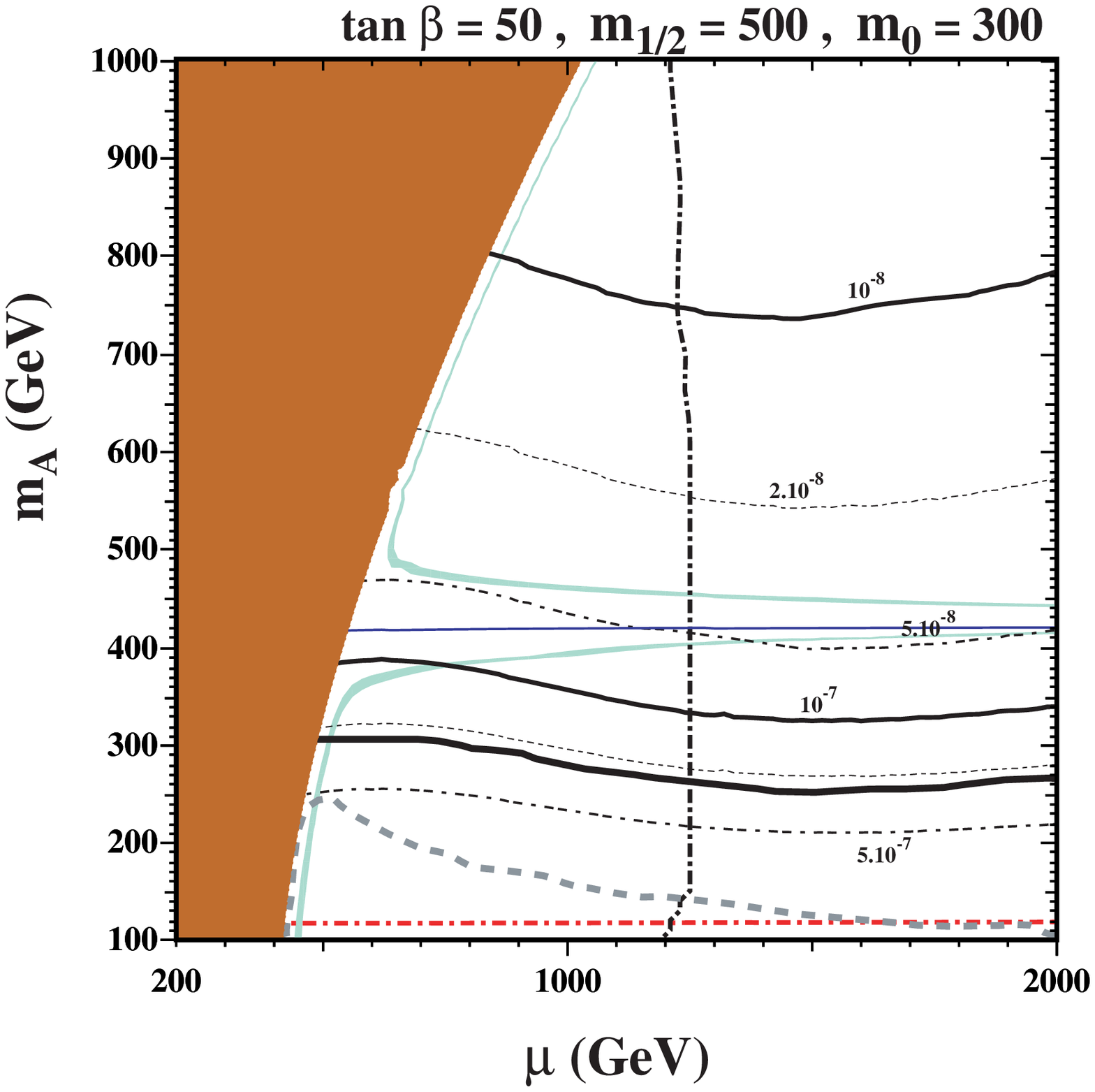}
\caption{\label{fig:bmm}
{\it Allowed regions in the $(\mu, M_A)$ planes for $m_{1/2} = 500$~GeV and 
$m_0 = 300$~GeV, for (a)  $\tan \beta = 40$, and (b) 
$\tan \beta = 50$. In each panel, the near-horizontal
solid blue line is the contour where $M_A = 2 m_\chi$, and the turquoise strips are 
those where the relic neutralino LSP density falls within the range 
favoured by WMAP and other cosmological and astrophysical observations. The 
LEP chargino limit is shown as a dashed black line and the GUT stability constraint as a dot-dashed black line. The regions disallowed because the ${\tilde \tau}_1$ 
would be the LSP are shaded brick-red.
Contours of the $\bmm$ branching ratio are labelled correspondingly, with the
current Tevatron limit the boldest black line, and the
CDMS constraint is shown as a dashed grey line.
There is no electroweak symmetry breaking in the polka-dotted region.}} 
\end{figure}


\begin{theacknowledgments}
I would like to thank J. Ellis, S. Heinemeyer, Y. Santoso, V. Spanos, and G. Weiglein
for many fruitful collaborations leading to the results summarized here. 
  This work was partially supported by DOE grant DE-FG02-94ER-40823. 
\end{theacknowledgments}



\bibliographystyle{aipprocl} 


\begin{thebibliography}{9}

\bibitem{vcmssm}
J.~R.~Ellis, K.~A.~Olive, Y.~Santoso and V.~C.~Spanos,
  Phys.\ Lett.\ B {\bf 573} (2003) 162
  [arXiv:hep-ph/0305212];
  J.~R.~Ellis, K.~A.~Olive, Y.~Santoso and V.~C.~Spanos,
  Phys.\ Rev.\ D {\bf 70} (2004) 055005
  [arXiv:hep-ph/0405110].
  
\bibitem{bfs}
R. Barbieri, S. Ferrara and C.A. Savoy, Phys.\ Lett. {\bf 119B} (1982) 343;
For reviews, see:
H.~P.~Nilles, Phys. Rep. {\bf 110} (1984) 1;
A.~Brignole, L.~E.~Ibanez and C.~Munoz,
arXiv:hep-ph/9707209,
published in {\it Perspectives on supersymmetry}, ed.
G.~L.~Kane, pp. 125-148. 

\bibitem{cmssmnew}
J.~R.~Ellis, G.~Ganis and K.~A.~Olive,
Phys.\ Lett.\ B {\bf 474} (2000) 314
[arXiv:hep-ph/9912324];
V.~D.~Barger and C.~Kao,
Phys.\ Lett.\ B {\bf 518} (2001) 117
[arXiv:hep-ph/0106189];
L.~Roszkowski, R.~Ruiz de Austri and T.~Nihei,
JHEP {\bf 0108} (2001) 024
[arXiv:hep-ph/0106334];
A.~B.~Lahanas and V.~C.~Spanos,
Eur.\ Phys.\ J.\ C {\bf 23} (2002) 185
[arXiv:hep-ph/0106345];
A.~Djouadi, M.~Drees and J.~L.~Kneur,
JHEP {\bf 0108} (2001) 055
[arXiv:hep-ph/0107316];
U.~Chattopadhyay, A.~Corsetti and P.~Nath,
Phys.\ Rev.\ D {\bf 66} (2002) 035003
[arXiv:hep-ph/0201001];
J.~R.~Ellis, K.~A.~Olive and Y.~Santoso,
New Jour.\ Phys.\  {\bf 4} (2002) 32
[arXiv:hep-ph/0202110];
H.~Baer, C.~Balazs, A.~Belyaev, J.~K.~Mizukoshi, X.~Tata and Y.~Wang,
JHEP {\bf 0207} (2002) 050
[arXiv:hep-ph/0205325];
R.~Arnowitt and B.~Dutta,
arXiv:hep-ph/0211417.


\bibitem{nonu} 
  M.~Olechowski and S.~Pokorski,
  Phys.\ Lett.\ B {\bf 344}, 201 (1995)
  [arXiv:hep-ph/9407404];
V.~Berezinsky, A.~Bottino, J.~Ellis, N.~Fornengo, 
               G.~Mignola and S.~Scopel,
               {\em Astropart.\ Phys.}  {\bf 5} (1996) 1, 
               hep-ph/9508249;
               M.~Drees, M.~Nojiri, D.~Roy and Y.~Yamada,
               {\em Phys.\ Rev.} {\bf D 56} (1997) 276, 
               [Erratum-ibid.\ {\bf D 64} (1997) 039901], 
               hep-ph/9701219;
               M.~Drees, Y.~Kim, M.~Nojiri, D.~Toya, K.~Hasuko and 
               T.~Kobayashi,
               {\em Phys.\ Rev.} {\bf D 63} (2001) 035008, 
               hep-ph/0007202;
               P.~Nath and R.~Arnowitt,
               {\em Phys.\ Rev.} {\bf D 56} (1997) 2820, 
               hep-ph/9701301;
               J.~R.~Ellis, T.~Falk, G.~Ganis, K.~A.~Olive and M.~Schmitt,
  Phys.\ Rev.\ D {\bf 58} (1998) 095002
  [arXiv:hep-ph/9801445];
J.~R.~Ellis, T.~Falk, G.~Ganis and K.~A.~Olive,
  Phys.\ Rev.\ D {\bf 62} (2000) 075010
  [arXiv:hep-ph/0004169];
               A.~Bottino, F.~Donato, N.~Fornengo and S.~Scopel,
               {\em Phys.\ Rev.} {\bf D 63} (2001) 125003, 
               hep-ph/0010203;
               S.~Profumo,
               {\em Phys.\ Rev.} {\bf D 68} (2003) 015006, 
               hep-ph/0304071;
               D.~Cerdeno and C.~Munoz,
               {\em JHEP} {\bf 0410} (2004) 015, 
               hep-ph/0405057;
               H.~Baer, A.~Mustafayev, S.~Profumo, A.~Belyaev and X.~Tata,
               {\em JHEP} {\bf 0507} (2005) 065, 
               hep-ph/0504001.


\bibitem{nuhm}
J.~Ellis, K.~Olive and Y.~Santoso,
\PL B~{\bf 539}, 107 (2002)
[arXiv:hep-ph/0204192];
J.~R.~Ellis, T.~Falk, K.~A.~Olive and Y.~Santoso,
Nucl.\ Phys.\ B {\bf 652}, 259 (2003)
[arXiv:hep-ph/0210205].


\bibitem{newBNL} [The Muon g-2 Collaboration],
                 {\it Phys. Rev. Lett.} {\bf 92} (2004) 161802, 
                 hep-ex/0401008;
                 G.~W.~Bennett {\it et al.}  [Muon G-2 Collaboration],
   ``Final report of the muon E821 anomalous magnetic moment measurement at
  Phys.\ Rev.\ D {\bf 73}, 072003 (2006)
  [arXiv:hep-ex/0602035].

\bibitem{g-2}
L.~L.~Everett, G.~L.~Kane, S.~Rigolin and L.~Wang,
Phys.\ Rev.\ Lett.\  {\bf 86} (2001) 3484 
[arXiv:hep-ph/0102145];
J.~L.~Feng and K.~T.~Matchev,
Phys.\ Rev.\ Lett.\  {\bf 86} (2001) 3480 
[arXiv:hep-ph/0102146];
E.~A.~Baltz and P.~Gondolo,   
Phys.\ Rev.\ Lett.\  {\bf 86} (2001) 5004 
[arXiv:hep-ph/0102147];
U.~Chattopadhyay and P.~Nath,
Phys.\ Rev.\ Lett.\  {\bf 86} (2001) 5854 
[arXiv:hep-ph/0102157];
S.~Komine, T.~Moroi and M.~Yamaguchi,
Phys.\ Lett.\ B {\bf 506} (2001) 93 
[arXiv:hep-ph/0102204];
J.~Ellis, D.~V.~Nanopoulos and K.~A.~Olive,
Phys.\ Lett.\ B {\bf 508} (2001) 65
[arXiv:hep-ph/0102331];
R.~Arnowitt, B.~Dutta, B.~Hu and Y.~Santoso,
Phys.\ Lett.\ B {\bf 505} (2001) 177  
[arXiv:hep-ph/0102344]
S.~P.~Martin and J.~D.~Wells,
Phys.\ Rev.\ D {\bf 64} (2001) 035003 
[arXiv:hep-ph/0103067];
H.~Baer, C.~Balazs, J.~Ferrandis and X.~Tata,
Phys.\ Rev.\ D {\bf 64} (2001) 035004 
[arXiv:hep-ph/0103280].


\bibitem{ehow3} J.~Ellis, S.~Heinemeyer, K.~Olive and G.~Weiglein,
                {\em JHEP} {\bf 0502} 013 (2005)  [arXiv:hep-ph/0411216].
                
                
\bibitem{ehow4} 
  J.~R.~Ellis, S.~Heinemeyer, K.~A.~Olive and G.~Weiglein,
  JHEP {\bf 0605}, 005 (2006)
  [arXiv:hep-ph/0602220].
                
\bibitem{wmap} D.~N.~Spergel {\it et al.},
  [arXiv:astro-ph/0603449].
  
  \bibitem{EHNOS}
J. Ellis, J.S. Hagelin, D.V. Nanopoulos, K.A. Olive
and M. Srednicki, Nucl. Phys. B {\bf 238} (1984) 453; see also
H. Goldberg, Phys. Rev. Lett. {\bf 50} (1983) 1419.

  
  
  \bibitem{eoss}
J.~R.~Ellis, K.~A.~Olive, Y.~Santoso and V.~C.~Spanos,
\PL {\bf B565} (2003) 176
[arXiv:hep-ph/0303043].

\bibitem{Baer}
H.~Baer and C.~Balazs,
{\it JCAP} {\bf 0305} (2003) 006
[arXiv:hep-ph/0303114];
A.~B.~Lahanas and D.~V.~Nanopoulos,
\PL {\bf B568} (2003) 55
[arXiv:hep-ph/0303130];
U.~Chattopadhyay, A.~Corsetti and P.~Nath,
\PR {\bf D68} (2003) 035005
[arXiv:hep-ph/0303201];
C.~Munoz,
{\it Int.\ J.\ Mod.\ Phys.\  } {\bf A19}, 3093 (2004)
[arXiv:hep-ph/0309346]
R.~Arnowitt, B.~Dutta and B.~Hu,
arXiv:hep-ph/0310103.

                        
                         \bibitem{stauco}
J.~R.~Ellis, T.~Falk and K.~A.~Olive,
Phys.\ Lett.\ B {\bf 444} (1998) 367
[arXiv:hep-ph/9810360];
J.~R.~Ellis, T.~Falk, K.~A.~Olive and M.~Srednicki,
Astropart.\ Phys.\  {\bf 13} (2000) 181
[Erratum-ibid.\  {\bf 15} (2001) 413]
[arXiv:hep-ph/9905481];
R.~Arnowitt, B.~Dutta and Y.~Santoso,
Nucl.\ Phys.\ B {\bf 606} (2001) 59
[arXiv:hep-ph/0102181];
M.~E.~G\'omez, G.~Lazarides and C.~Pallis,
Phys. Rev. D {\bf D61} (2000) 123512
[arXiv:hep-ph/9907261];
  Phys.\ Lett. {\bf B487} (2000) 313
[arXiv:hep-ph/0004028];
  Nucl. Phys. B {\bf B638} (2002) 165
[arXiv:hep-ph/0203131];
T.~Nihei, L.~Roszkowski and R.~Ruiz de Austri,
  JHEP {\bf 0207} (2002) 024
[arXiv:hep-ph/0206266].


\bibitem{bench}
 M.~Battaglia, A.~De Roeck, J.~Ellis, F.~Gianotti, 
                 K.~Olive and L.~Pape,
                 {\em Eur. Phys. J.} {\bf C 33} (2004) 273, 
                 hep-ph/0306219;

  \bibitem{otherAnalyses} 
                                          H.~Baer, A.~Belyaev, T.~Krupovnickas and X.~Tata,
                        {\em JHEP} {\bf 0402} (2004) 007, 
                        hep-ph/0311351;
  J.~R.~Ellis, K.~A.~Olive, Y.~Santoso and V.~C.~Spanos,
  Phys.\ Lett.\ B {\bf 603}, 51 (2004)
  [arXiv:hep-ph/0408118].
                B.~Allanach, G.~Belanger, F.~Boudjema and A.~Pukhov,
                        hep-ph/0410091.
                        
                        
\bibitem{funnel}
H.~Baer and M.~Brhlik,
Phys.\ Rev.\ D {\bf 53} (1996) 597
[arXiv:hep-ph/9508321];
H.~Baer, M.~Brhlik, M.~A.~Diaz, J.~Ferrandis, P.~Mercadante,  
P.~Quintana and X.~Tata,
Phys.\ Rev.\ D {\bf 63} (2001) 015007
[arXiv:hep-ph/0005027];
A.~B.~Lahanas and V.~C.~Spanos,
Eur.\ Phys.\ J.\ C {\bf 23} (2002) 185
[arXiv:hep-ph/0106345].

\bibitem{efgosi}
J.~R.~Ellis, T.~Falk, G.~Ganis, K.~A.~Olive and M.~Srednicki,
Phys.\ Lett.\ B {\bf 510} (2001) 236
[arXiv:hep-ph/0102098].

  \bibitem{Dedes}
  A.~Dedes, H.~K.~Dreiner and U.~Nierste,
  Phys.\ Rev.\ Lett.\  {\bf 87} (2001) 251804
  [arXiv:hep-ph/0108037];
  R.~Arnowitt, B.~Dutta, T.~Kamon and M.~Tanaka,
  Phys.\ Lett.\ B {\bf 538} (2002) 121
  [arXiv:hep-ph/0203069];
      S.~Baek, P.~Ko and W.~Y.~Song,
  Phys.\ Rev.\ Lett.\  {\bf 89} (2002) 271801
  [arXiv:hep-ph/0205259];
  C.~S.~Huang and X.~H.~Wu,
  Nucl.\ Phys.\ B {\bf 657} (2003) 304
  [arXiv:hep-ph/0212220].
  S.~Baek, Y.~G.~Kim and P.~Ko,
  JHEP {\bf 0502} (2005) 067
  [arXiv:hep-ph/0406033];
  H.~Baer, C.~Balazs, A.~Belyaev, J.~K.~Mizukoshi, X.~Tata and Y.~Wang,
  JHEP {\bf 0207} (2002) 050
  [arXiv:hep-ph/0205325];
  E.~Lunghi, W.~Porod and O.~Vives,
  arXiv:hep-ph/0605177.
                        
  \bibitem{bmm}
J.~R.~Ellis, K.~A.~Olive and V.~C.~Spanos,
  Phys.\ Lett.\ B {\bf 624} (2005) 47
  [arXiv:hep-ph/0504196].

\bibitem{bmm2}
  J.~R.~Ellis, K.~A.~Olive, Y.~Santoso and V.~C.~Spanos,
  JHEP {\bf 0605}, 063 (2006)
  [arXiv:hep-ph/0603136].

 \bibitem{cdf} 
F.~Abe {\it et al.}  [CDF Collaboration],
  Phys.\ Rev.\ D {\bf 57} (1998) 3811;
   D.~Acosta {\it et al.}  [CDF Collaboration],
  Phys.\ Rev.\ Lett.\  {\bf 93} (2004) 032001
  [arXiv:hep-ex/0403032];
 V.~M.~Abazov {\it et al.}  [D0 Collaboration],
  Phys.\ Rev.\ Lett.\  {\bf 94}, 071802 (2005)
  [arXiv:hep-ex/0410039];
  The D0 Collaboration, D0note, 4733-CONF;
  http://www-d0.fnal.gov/Run2Physics/WWW/results/prelim/B/B21/B21.pdf;
M. Herndon, The CDF and D0 Collaborations, 
FERMILAB-CONF-04-391-E. Published Proceedings 32nd International 
Conference on High-Energy Physics (ICHEP 04), Beijing, China, August 16-22, 2004;
The CDF Collaboration, CDF note 7670;
http://www-cdf.fnal.gov/physics/new/bottom/050407.blessed-bsmumu/.

\bibitem{cdfnew}
The CDF Collaboration, CDF Public Note 8176;
http://www-cdf.fnal.gov/physics/new/bottom/060316.blessed-bsmumu3/


\bibitem{fp}
J.~L.~Feng, K.~T.~Matchev and T.~Moroi,
{\it Phys.\ Rev.\ D} {\bf 61} (2000) 075005
[arXiv:hep-ph/9909334].



  \bibitem{Davier}
M.~Davier, S.~Eidelman, A.~H\"ocker and Z.~Zhang,
               {\it Eur.\ Phys.\ J.}\  {\bf C 31} (2003) 503,
               hep-ph/0308213;
                see also
K.~Hagiwara, A.~Martin, D.~Nomura and T.~Teubner,
                  {\it Phys. Rev.} {\bf D 69} (2004) 093003, 
                  hep-ph/0312250;
S.~Ghozzi and F.~Jegerlehner,
                      {\it Phys. Lett.} {\bf B 583} (2004) 222,
                      hep-ph/0310181;
 M.~Knecht,
                     hep-ph/0307239;
K.~Melnikov and A.~Vainshtein,
\PR {\bf D70} (2004) 113006
[arXiv:hep-ph/0312226].
 J.~de Troconiz and F.~Yndurain,
                   hep-ph/0402285;
  M.~Passera,
  Nucl.\ Phys.\ Proc.\ Suppl.\  {\bf 155}, 365 (2006)
  [arXiv:hep-ph/0509372].
  
\bibitem{eoss8}
J.~R.~Ellis, K.~A.~Olive, Y.~Santoso and V.~C.~Spanos,
  Phys.\ Rev.\ D {\bf 71}, 095007 (2005)
  [arXiv:hep-ph/0502001].

\bibitem{SVZ}
M.\ A.\ Shifman, A.\ I.\ Vainshtein and V.\ I.\ Zakharov, 
{\em Phys.\ Lett.} {\bf 78B}, 443 (1978); \\ A.\ I.\ Vainshtein,
V.\ I.\ Zakharov and M.\ A.\ Shifman, {\em Usp.\ Fiz.\ Nauk} 
{\bf 130}, 537 (1980).

\bibitem{gdm} J.~Ellis, K.~Olive, Y.~Santoso and V.~Spanos,
              {\em Phys.\ Lett.} {\bf B 588} (2004) 7, 
              hep-ph/0312262;
J.~Ellis, K.~Olive and E.~Vangioni,
              {\em Phys.\ Lett.} {\bf B 619} (2005) 30, 
              astro-ph/0503023;
J.~Feng, S.~Su and F.~Takayama,
               {\em Phys.\ Rev.} {\bf D 70} (2004) 075019, 
               hep-ph/0404231;
               {\em Phys.\ Rev.} {\bf D 70} (2004) 063514, 
               hep-ph/0404198;
               J.~Feng, A.~Rajaraman and F.~Takayama,
               {\em Phys.\ Rev.\ Lett.} {\bf 91} (2003) 011302, 
               hep-ph/0302215.

\bibitem{Moortgat} K.~Hamaguchi, Y.~Kuno, T.~Nakaya and M.~Nojiri,
                   {\em Phys.\ Rev.} {\bf D 70} (2004) 115007, 
                   hep-ph/0409248;
                   J.~Feng and B.~Smith,
                   {\em Phys.\ Rev.} {\bf D 71} (2005) 015004
                   [Erratum-ibid.\ {\bf D 71} (2005) 0109904],
                   hep-ph/0409278;
                   A.~De Roeck, J.~Ellis, F.~Gianotti, F.~Moortgat, 
                   K.~Olive and L.~Pape,
                   hep-ph/0508198.
                   
                   
                   
\bibitem{oldmt} V.~Abazov {\it et al.} [D0 Collaboration],
                {\em Nature} {\bf 429} (2004) 638, 
                hep-ex/0406031;\\
                P.~Azzi {\it et al.}
                [CDF Collaboration, D0 Collaboration],
                hep-ex/0404010.
                
                
\bibitem{newmt} CDF Collaboration, D0 Collaboration and Tevatron Electroweak 
                Working Group, hep-ex/0507091.
               
\bibitem{newestmt} Tevatron Electroweak Working Group,
                   hep-ex/0603039.
\bibitem{evennewstmt}
 T.~E.~W.~Group,
  arXiv:hep-ex/0608032.
  
  
  \bibitem{ehow4p}
    J.~Ellis, S.~Heinemeyer, K.~A.~Olive and G.~Weiglein,
  arXiv:hep-ph/0604180.
  
  \bibitem{LEPHiggsSM} LEP Higgs working group,
                     {\em Phys. Lett.} {\bf B 565} (2003) 61,
                     hep-ex/0306033;
ALEPH, DELPHI, L3, OPAL
                       Collaborations and LEP Working Group for Higgs
                       boson searches, 
                       hep-ex/0602042.
                       
                       \bibitem{sps}
                        B.~Allanach et al.,
              {\em Eur. Phys. J.} {\bf C 25} (2002) 113, 
              hep-ph/0202233;


\bibitem{lhcilc} G.~Weiglein {\it et al.} [LHC / ILC Study Group],
                 {\em Phys. Rept.} {\bf 426} (2006) 47,
                 hep-ph/0410364.


\bibitem{cdms}
  D.~S.~Akerib {\it et al.}  [CDMS Collaboration],
   ``Limits on spin-independent WIMP nucleon interactions from the two-tower
  Phys.\ Rev.\ Lett.\  {\bf 96}, 011302 (2006)
  [arXiv:astro-ph/0509259].
  
  

 

\end{thebibliography}


\end{document}